\documentclass[fleqn,10pt]{natclass}
\title{Phase and amplitude imaging with quantum correlations through Fourier Ptychography}

\usepackage{bm}
\usepackage{textcomp}
\usepackage{float}
\usepackage{xcolor}

\author[1]{Tomas Aidukas}
\author[1]{Pavan Chandra Konda}
\author[1]{Andrew R. Harvey}
\author[1]{Miles J. Padgett}
\author[1,*]{Paul-Antoine Moreau}

\affil[1]{School of Physics and Astronomy, University of Glasgow, G12 8QQ, UK}

\affil[*]{paul-antoine.moreau@glasgow.ac.uk}

\usepackage{soul}
\soulregister\cite7
\soulregister\ref7
\soulregister\pageref7

\begin{abstract}
Extracting as much information as possible about an object when probing with a limited number of photons is an important goal with applications from biology and security to metrology. Imaging with a few photons is a challenging task as the detector noise and stray light are then predominant, which precludes the use of conventional imaging methods. Quantum correlations between photon pairs has been exploited in a so called 'heralded imaging scheme' to eliminate this problem. However these implementations have so-far been limited to intensity imaging and the crucial phase information is lost in these methods. In this work, we propose a novel quantum-correlation enabled Fourier Ptychography technique, to capture high-resolution amplitude and phase images with a few photons. This is enabled by the heralding of single photons combined with Fourier ptychographic reconstruction. We provide experimental validation and discuss the advantages of our technique that include the possibility of reaching a higher signal to noise ratio and non-scanning Fourier Ptychographic acquisition.
\end{abstract}

\begin{document}

\flushbottom
\maketitle

\thispagestyle{empty}

\noindent 

\section*{Context and objectives}

Imaging with low-light-levels is central to a large range of applications from biology and security to high precision metrology. In particular, imaging delicate objects such as biological cells or photosensitive chemicals that requires low light illumination to preserve their integrity. The ultimate goal for such applications is to recover as much image information as possible per photon. For that purpose, extracting both phase and intensity information in low-light imaging is a crucial aim. Indeed in some circumstances phase contains important information about an object that is not present in the intensity information alone. For example, most cells are effectively transparent, but their shape can still be inferred from the phase information\cite{tian2015computational}.\\

Traditional imaging schemes suffer from noise either due to the sensor noise or due to parasitic light from the surroundings. In very low light regimes these source of noise become predominant leading to images presenting poor signal to noise ratio (SNR). The use of quantum imaging in this context allows the improvement of the images quality. For a review on quantum imaging techniques and the kind of improvement they allow see ref\cite{moreau2019review}. In particular, a way to remedy the aforementioned noise issue is to implement a particular quantum imaging scheme called 'heralded imaging scheme' \cite{aspden2013epr,morris2015imaging}. In heralded imaging (HI) a heralding detector records one photon from a photon pair and triggers the imaging detector to record exactly when the second photon is incident on the sensor. The imaging detector, usually an ICCD camera, is gated for a few picoseconds, and records only the photons originating from a pair. This therefore eliminates most of the undesired events such as background light and dark counts from camera since they are not correlated with heralding events. The use of such techniques to reach very low intensities were so far limited to intensity only imaging or qualitative phase contrast only imaging ~\cite{aspden2016heralded,lu2015quantum}, limiting therefore the amount of information extracted about an object. Here we demonstrate that quantitative phase-amplitude imaging with a small number of photons is possible by using a Fourier ptychographic acquisition technique enabled by quantum correlations through a heralded imaging scheme.\\

\begin{figure*}
\center
\includegraphics[width=0.85
\linewidth]{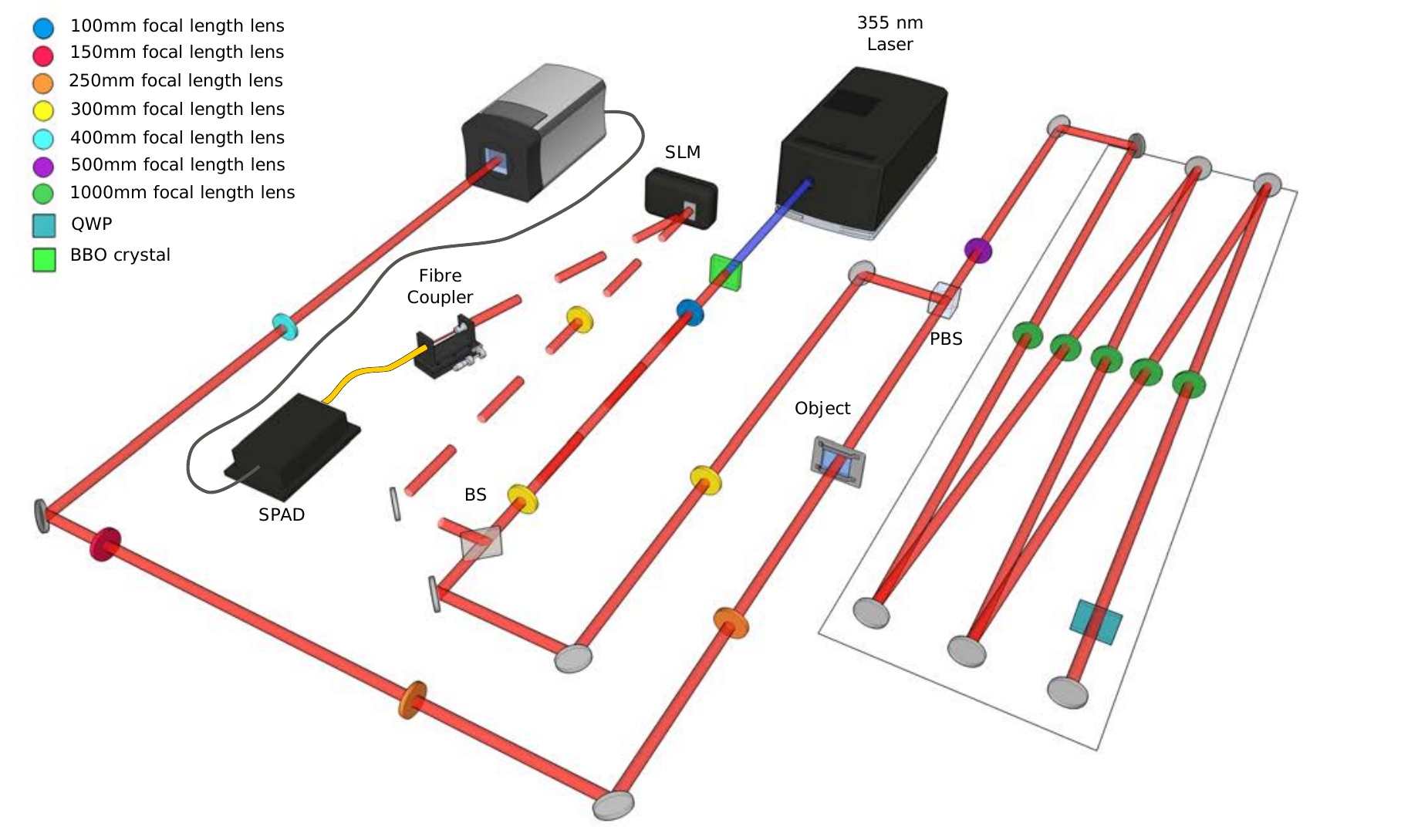}
\caption{\textbf{Schematics of the experimental setup.} A 355 nm laser pumps a BBO crystal to produce collinear downconverted photon pairs at 710 nm. The signal and idler photons (respectively represented in plain and dashed lines) are separated using a beam splitter. The far-field of the crystal is imaged onto the object and the ICCD camera in the other arm. The image plane of the crystal is imaged in the fiber plane in the idler arm. The relative displacement of the fiber within the the idler beam is produced by an SLM placed in the Far-Field of the crystal in the heralding arm. The ICCD camera is then triggered by the SPAD to obtain HI coherent images. An image-preserving delay line is introduced to compensate for the electronic delays in triggering the camera (delay line shown within the rectangle on the right part of the setup). An aperture can be introduced at the object focal plane of the last lens in front of the camera to fix the numerical aperture of the imaging system.}
\label{fig:setup}
\end{figure*}

Fourier ptychography (FP) is a synthetic aperture technique which uses low-resolution objectives and tilt-shift illumination \cite{wicker2014resolving} to synthesize high-resolution, amplitude and phase images \cite{zheng2013wide}. It replaces the traditional interferometric methods \cite{tippie2011high} with a computational iterative algorithm \cite{zheng2013wide}. This simplifies the experimental setup, enabling for example the development of 3D printed low-cost microscopes \cite{konda2017miniature,konda2018parallelized}. FP has notedly been used to demonstrate quantitative phase imaging via Fourier ptychographic microscopy\cite{ou2013quantitative}, 3D refocusing\cite{dong2014aperture}, and highly-resolved in vitro microscopy through the acquisition of gigapixel-scale phase and amplitude images\cite{tian2015computational}. For a review on Ptychographic image acquisition technique see ref~\cite{rodenburg2008ptychography}.\\

In FP, a plane wave is used to coherently illuminate an object. By changing the angle of this plane wave illumination (with respect to the object), different images can be recorded. A wide range of illumination angles are generally used to capture several images, which are then combined using FP reconstruction algorithms. As we will show bellow, Fourier ptychography can be combined with quantum correlations to obtain low photon number phase and intensity images. Our work is enabled by using parametric down-conversion as the illumination source which exhibits Einstein-Podolsky-Rosen (EPR) correlations between signal and idler photons. A pump beam of sufficiently large diameter is used such that the down-coverted twin photons are created over a wide range of different spatial modes, resulting in an extended illumination source, thereby allowing the acquisition of various images of the same object corresponding to different illumination directions without having to change the actual illumination conditions of the object. The mechanism allowing such an acquisition uses the transverse spatial correlations exhibited by the two particles. When the momentum of one of the particles is measured with a precise value, the other particle is projected onto a state with a precisely defined opposed momentum. Therefore, by post-selecting in an image the photons whose twin photon were detected with a given momentum, one ensures that all the post selected photons had the same state corresponding to a plane wave when incident on the imaged object. This ensures that such post-selected images will virtually be that of the object acquired under a particular illumination direction. Post-selecting images on different momentum measurement outcome for the twin photon allows the acquisition of images virtually illuminated from many distinct directions. These images are combined together using an iterative FP algorithm to synthesise a wide-field of view, high-resolution amplitude and phase image. In the experimental demonstration reported here, the acquired images exhibit Poissonian noise due to the low number of photons in each image pixel. To minimise the impact of this Poisonian noise on the reconstruction we use a sequential Gauss-Newton optimisation method to perform the reconstruction. Such an optimisation method has previously been proven to be robust in the presence of Poissonian noise in the context of Fourier ptychography\cite{yeh2015experimental}.\\

The illumination source used in the present realisation are photons generated through the spontaneous parametric down-conversion (SPDC) process. This type of source has been used as illumination in several low-light level applications and techniques\cite{strekalov1995observation,walborn2006quantum,dixon2012quantum}. In particular SPDC has been used as a way to generate the optical manifestation of the EPR state in its original domain i.e. in the spatial variables of position and momentum \cite{howell2004realization,moreau2012realization,edgar2012imaging,moreau2014einstein}. These EPR states have been used in single-photon imaging, for example to implement quantum ghost imaging\cite{pittman1995optical,aspden2013epr,morris2015imaging}, which can be extended to trans-wavelength imaging schemes\cite{aspden2015photon}. SPDC can also be used to record video of single-photon double-slit interference\cite{aspden2016video} and we used it recently to compare the resolution limits of quantum ghost imaging to classical imaging\cite{moreau2018resolution}. EPR states have also been used to perform quantum imaging with undetected photons\cite{lemos2014quantum} and importantly for the presently reported realisation to implement heralded imaging~\cite{morris2015imaging}.

The nature of SPDC is that the signal and idler photons exhibits correlations in their transverse position or momentum, this results in peculiar features concerning its optical spatial coherence. Illuminating an object with one of the SPDC beams and detecting the intensity of that beam gives an incoherent imaging scheme. It was shown that by detecting coincidences between the signal and idler SPDC beams, by using on one beam a single mode detector, leads to retrieval of the underlying coherence of the state\cite{tasca2013influence}. In the present work this coherence is needed as the Fourier Ptychography method requires the acquisition of a spatially coherent image. We show in the following section how correlations generated with an SPDC source can be used to perform Fourier ptychography.

\begin{figure*}
\centering
\includegraphics[width=\linewidth]{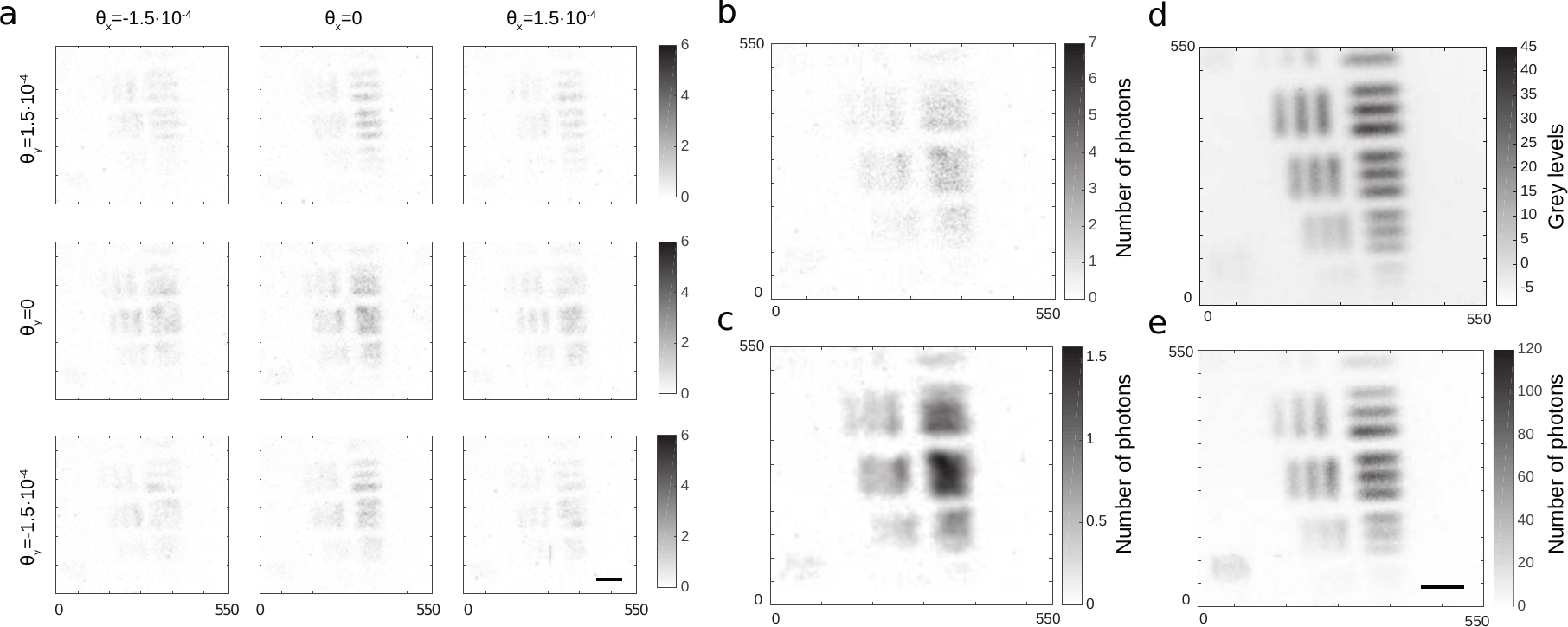}
\caption{\textbf{Acquired and reconstructed images of a test target.} (\textbf{a}) A set of images acquired with different illumination angles $(\theta_x,\theta_y)$ by post-selecting on different idler positions in an HI acquisition. Color bars indicate the number of photons. (\textbf{b}) Image acquired with normal illumination (0,0) acquired through HI, resulting in a spatially coherent imaging. (\textbf{c}) Same image regularized with an optimized regularization factor $\lambda=1.69$, see \cite{morris2015imaging} for a description of the regularization method. (\textbf{d}) Image of the test target acquired through DI, resulting in an incoherent imaging. (\textbf{e}) Simulation of the incoherent imaging by simply summing the HI images acquired for an ensemble of different illumination angles. N.b. the contrast has been inverted to make the photons within the sparse images more prominant. Scale bar, 400 \textmu{}m.}
\label{fig:imgs}
\end{figure*}

\section*{Results}

\subsection*{Principle of the acquisition.}
We show here that it is possible to use both the spatial coherence, manifest through coincident detection, and the SPDC spatial extent to realize a Fourier ptychographic reconstruction which requires the spatially coherent illumination of object from a range of different directions. For a classical source, its extended nature would lead to a loss of this spatial coherence.\\

Let us consider an EPR state that can be written in the following way:
\begin{equation}
\left| EPR \right \rangle\propto\int_{-\infty}^{\infty}\left|x\right \rangle_s\left|x \right \rangle_i dx\propto\int_{-\infty}^{\infty}\left|k\right \rangle_s\left|-k \right \rangle_i dk
\end{equation}
Where we use the proportionality to avoid tracking normalisation coefficients and $i$ ($s$) stand for signal (idler). The associated density matrix has the following form :
\begin{eqnarray}
\boldsymbol\rho_{EPR}&\propto&\iint_{-\infty}^{\infty}\left|x\right\rangle_s\left|x \right \rangle_i\left\langle x'\right|_s\left\langle x'\right|_i dx dx'\\&\propto&\iint_{-\infty}^{\infty}\left|k\right\rangle_s\left|-k \right \rangle_i\left\langle k'\right|_s\left\langle -k'\right|_i dk dk'
\end{eqnarray}
The state of the signal alone can be retrieved by tracing this operator on the idler:
\begin{equation}
\boldsymbol\rho_{s}=\mathrm{tr}_i\left(\boldsymbol\rho_{EPR}\right)\propto\int_{-\infty}^{\infty} \left|k \right \rangle_s\left\langle k\right|_sdk
\end{equation}
which is a statistical mixture state of plane waves i.e. a spatially incoherent illumination. Now if one considers the following measurement operator corresponding to the detection of the idler photon with a given wave vector noted $k_f$,
\begin{equation}
\mathbf{P}=\mathbb{I}_s\otimes\left| k_f \right\rangle_i\left\langle k_f \right|_i 
\end{equation}
After projection and tracing out the state on the idler, one can show that the state of the signal becomes, 
\begin{equation}
\boldsymbol\rho_{s}=\mathrm{tr}_i\left(\mathbf{P}\boldsymbol\rho_{EPR}\mathbf{P}\right)\propto\int_{-\infty}^{\infty} e^{-ik_fx}\left|x \right \rangle_s\left\langle x'\right|_se^{ik_fx'}dxdx'
\end{equation}
corresponding to a pure state of the form,
\begin{equation}
\left| \Psi_s \right\rangle = \left| k_f \right\rangle \propto \int_{-\infty}^{\infty} e^{-ik_fx}\left| x \right \rangle dx
\end{equation}
This is a plane wave whose tilt compared to the optical axis is parametrised by the position of detection of the idler photon in the wave vector plane $k_f$. Equivalently if the idler position is detected at the position $x_f$ then the state of the signal photon becomes:
\begin{equation}
\left| \Psi_s \right\rangle = \left| x_f \right\rangle \propto \int_{-\infty}^{\infty} e^{ix_fk}\left| k \right \rangle dk
\label{illumstate}
\end{equation}
Such a state corresponds to an inclined plane wave incident on an object placed in the far-field of the crystal. Such wave planes are exactly the illumination that is needed to perform a Fourier ptychographic (FP) reconstruction. By illuminating an object with an SPDC signal beam and detecting correlations between the momentum of the idler photons (defined by the transverse position of the single-mode detector) and the positions of the correlated signal photons (after the object) one can, in principle, record a complete Fourier ptychographic reconstruction without changing the illumination of the object. A proposal to harness EPR-like correlations in a similar way was recently proposed in the context of plenoptic imaging~\cite{pepe2016correlation}.\\

Our aim in the following section is to experimentally demonstrate such a protocol, and to show that one can use this scheme to perform a phase and intensity reconstruction of an object with images acquired with a low number of photons per image pixel.

\subsection*{Experimental implementation}
The setup used for the experimental demonstration is presented in Fig. \ref{fig:setup}. The EPR source is formed through spontaneous parametric down-conversion of a UV Laser. We use an ICCD camera triggered by a single-photon avalanche photo-diode (SPAD) to detect images of heralded single photons. The photons detected by the SPAD are first collected with a single-mode fiber, which ensures that the detected idler photons are collected at a given position $(x_f,y_f)$ inside the beam. We therefore post-select the results for a certain position $(x_f,y_f)$ for the idler photons $i$, which is simply done by triggering the ICCD camera conditionally on the detection of a photon by the SPAD. This triggering ensures that the acquired images  correspond to images obtained by illuminating the object with the state described in equation (\ref{illumstate}). We then change the position of the fiber within the spatial extent of the idler beam $(x_f,y_f)$ by changing the period of the blazed phase grating displayed on an SLM, thereby changing the deflection angle of the idler beam. A set of images is acquired, each image corresponding to different values of $(x_f,y_f)$ and therefore to different illumination angles $(\theta_x,\theta_y)$ of the object. By recording many such heralded images, One can then perform a complete Fourier Ptychographic acquisition.\\

The plane wave synthesis in our system can be explained using Klyshko advanced wave picture \cite{klyshko1988simple,aspden2014experimental}. In this picture the heralding detector is replaced by a classical light source contra-propagating in the heralding arm up to the crystal that then acts simply as a mirror reflecting the light through the rest of the system to the camera. The light originating from a single point on the heralding detector side forms a collimated beam in the object plane in the imaging arm. The single mode fiber, defines a single point in the heralding arm, resulting in a single plane wave in the object plane to form a spatially coherent image on the camera. The position of this single mode fiber determines the angle of illumination and the spatial frequencies sampled. Figure 2\textbf{a} shows few of images of an object acquired for various periods of blazed gratings.\\

\begin{figure}[ht]
\centering
\includegraphics[width=\linewidth]{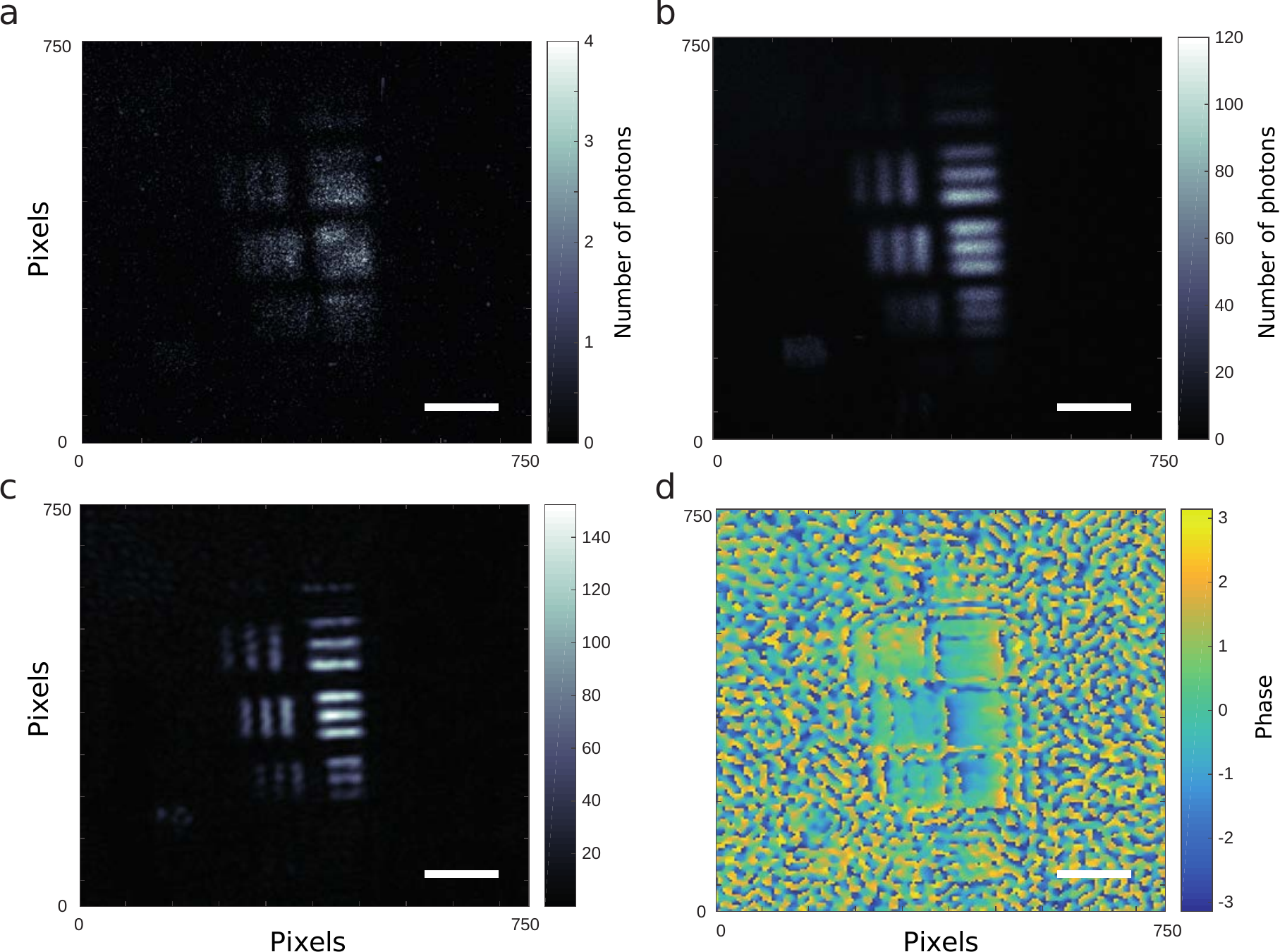}
\caption{\textbf{Fourier Ptychographic reconstruction of a portion in a USAF resolution test chart.} (\textbf{a}) Image acquired with normal illumination. (\textbf{b}) Incoherent sum of the 225 images acquired for the acquisition. (\textbf{c}) Intensity reconstruction of the USAF target through FP. (\textbf{d}) Phase reconstruction of the USAF target through FP. Scale bar, 600 \textmu{}m.}
\label{fig:Target2}
\end{figure}

\begin{figure}[ht]
\centering
\includegraphics[width=\linewidth]{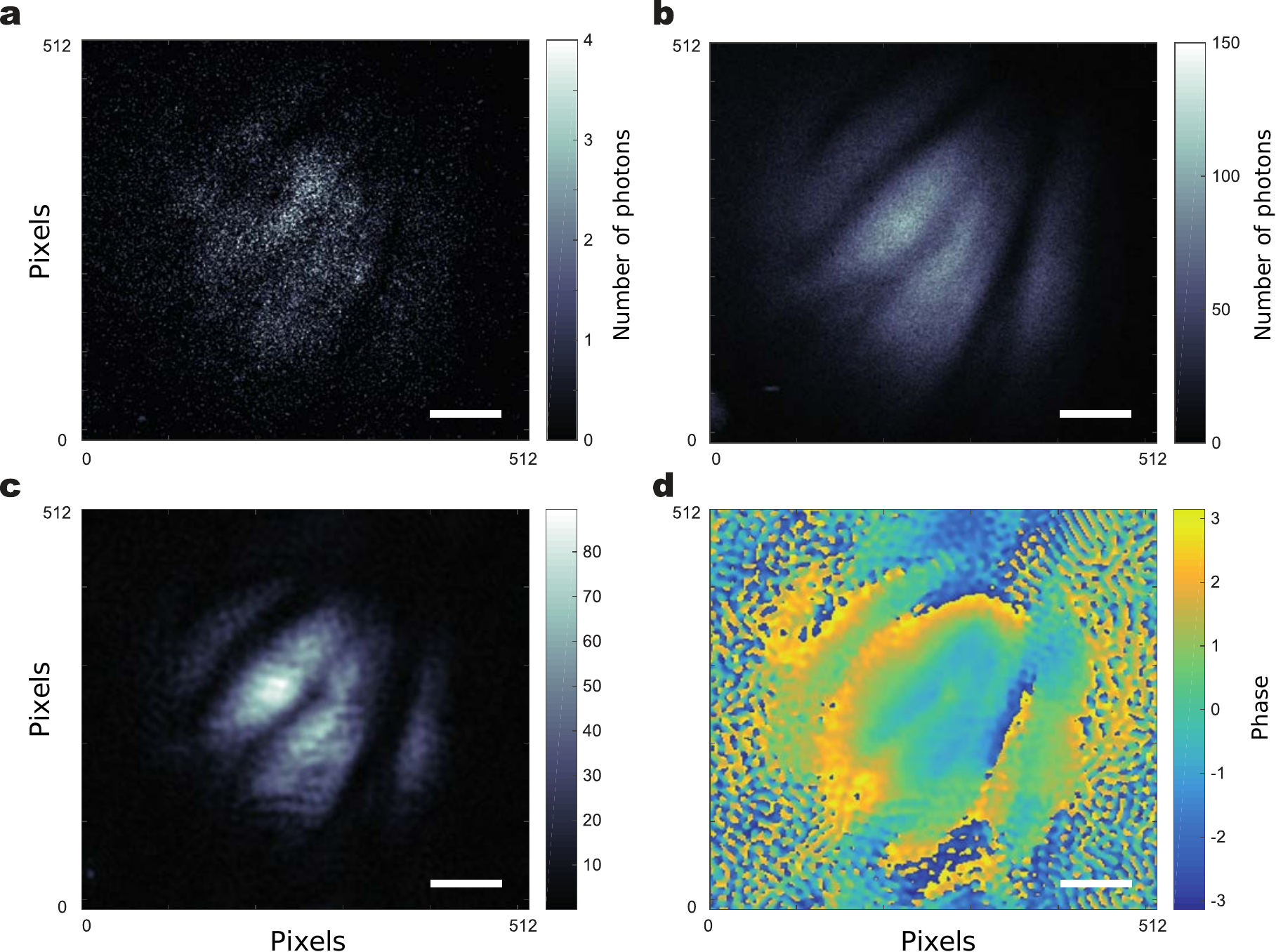}
\caption{\textbf{Fourier Ptychographic reconstruction of a portion of wasp wing.} (\textbf{a}) (\textbf{a}) Image acquired with normal illumination. (\textbf{b}) Incoherent sum of the 225 images acquired for the acquisition. (\textbf{c}) Intensity reconstruction of the wasp wing through FP. (\textbf{d}) Phase reconstruction of the wasp wing through FP. Scale bar, 400 \textmu{}m.}
\label{fig:LowNA}
\end{figure}

With this experimental setup we can acquire the images in two different ways. The first method is the heralded imaging (HI) described above, through which we perform the Ptychographic acquisition. In this case, the camera is triggered by the external trigger pulses corresponding to detection events of idler photons. But for comparison, we can also acquire images through direct imaging (DI) by changing only the way the camera is triggered. In this later case the camera is triggered using its internal trigger mechanism resulting in images generated by a random subset of photons that pass through the object and arrive at the camera sensor during the camera trigger window. Since either beam of the SPDC light is spatially incoherent when considered independently from the other, the DI image of the object that is obtained is an image resulting from spatially incoherent illumination. Fig. \ref{fig:imgs} compares the image acquired with the HI method (\textbf{a}, \textbf{b} and \textbf{c}) and the one acquired with DI (\textbf{d}).\\
The images obtained through the DI method are acquired in a continuous trigger mode and we use the camera detector as a conventional linear detector. We acquire the DI images by accumulating frames that are not photon sparse, and due to the long integration time have a high background that we must subsequently, numerically, subtract. In contrast, the images acquired through heralded imaging are composed of photo detection events and do not need to be background subtracted. This is visible in Fig. \ref{fig:imgs}, the colorbar in panel \textbf{d} goes bellow 0 due to the supplemental noise present in the background and the subtraction of the mean background. This is not the case in the other panels n Fig. \ref{fig:imgs} for which no background substraction have to be performed. A full comparison between heralded, ghost and direct imaging is out of the scope of the present article, for a discussion and comparison of the performances of heralded and direct imaging see \cite{morris2015imaging}\\
One can observe in Fig. \ref{fig:imgs} how switching from HI to DI result in the loss of the spatial coherence. Indeed one can see more details in the images acquired through DI and background subtracted Fig. \ref{fig:imgs}\textbf{d} compared to the one acquired through the HI Fig. \ref{fig:imgs}\textbf{b} and \textbf{c}. Moreover, one can see in Fig. \ref{fig:imgs} that by summing the detected intensity of the the different HI acquisitions for different equivalent illumination angles (see Fig. \ref{fig:imgs}\textbf{e}), one retrieves an image that looks very similar to the DI image with the difference that no background subtraction has had to be performed in the HI case. The comparison is consistent with the notion that the DI image is a equivalent to of an incoherent sum of the HI images i.e. a spatially incoherent imaging scheme.\\
However, because in our scheme the individual HI images are accessible, it allows us to perform Fourier Ptychographic reconstructions with very low photon numbers as exemplified by the fact that no background subtraction has to be performed on any of the images. As dicussed above, this is done by triggering the camera with a very short gate-time of only a few nanoseconds synchronised to when a single photon event is likely to be incident on the sensor. By doing this one removes the detector dark noise that would occur when using the long exposure time normally required for low light imaging. This protocol allows us to acquire images with a very high signal to noise ratio relative to the low number of detected events. It should be noted that this present experimental demonstration of principle presents a low efficiency in terms of photon detected compared to the number of photons illuminating the object (about 1\%), however, our proposed scheme could potentially lead to single snapshot Fourier ptychography with a very weak illumination of the sample by replacing the single mode detector with a second camera, the coincident HI image associated with each pixel of which would correspond to a different illumination angle of the object.\\

\subsection*{Reconstructed images}
The heralded images acquired as described above can be processed using a Fourier ptychography reconstruction algorithm. It is based on the alternate projection type algorithm developed by Gerchberg-Saxton~\cite{gerchberg1972practical} and later improved by Fienup \cite{ fienup1987reconstruction}. In these algorithms the amplitude of the complex field is known in one plane and a support structure for the amplitude and phase is known in the other plane. In an imaging system the support structure is the lens pupil shape, i.e., all the pixel values outside the lens pupil area should be zero. During the reconstruction, a random value is assigned for the unknown phase in the first plane and the complex field is propagated to the second plane. In this plane all the values outside the lens pupil area are set to zero and this distribution is then back propagated to the first plane. The amplitudes in the new complex field will be replaced by the known amplitudes from the experiment. These back and forth propagations are performed several times until the iterative algorithm finds the right phase.\\
In FP, the complex field is propagated between the Fourier plane and the image plane. Here, the alternate projection is performed for every illumination angle in sequence to complete one iteration, where the pupil position (support constraint) depends on the illumination angle. Each illumination angle updates a unique set of spatial frequencies in the object’s Fourier plane, hence by combining all the illuminations a large synthetic aperture is formed. In these datasets, an overlap between the spatial frequencies sampled by adjacent illumination angles is maintained to ensure the convergence \cite{horstmeyer2014overlapped}. This overalp redundancy deals system errors such as misalignment, aberrations and noise \cite{ou2014embedded,konda2015high}. In this work the images are subject to photon shot-noise arising since the photon counts are extremely low, around 1 photon per pixel (see Fig. \ref{fig:imgs}(c)), resulting in low signal to noise ratio. We implement a second order quasi Gauss-Newton based method for updating spatial frequencies in the Fourier plane \cite{yeh2015experimental}. This combined with an adaptive step-size approach \cite{zuo2016adaptive} provides a robust convergence of our data.\\

In Fig. \ref{fig:Target2} a USAF resolution test target is imaged to demonstrate the resolution improvement quantitatively and in \ref{fig:LowNA} a part of a wasp wing is imaged to demonstrate the resolution improvement and the phase recovery. In both cases we show that we have been able to obtain a reconstruction of the object in both intensity (Fig. \ref{fig:Target2}\textbf{c} and \ref{fig:LowNA}\textbf{c}) and phase  (Fig. \ref{fig:Target2}\textbf{d} and \ref{fig:LowNA}\textbf{d}). We also show for comparison in Fig. \ref{fig:Target2}\textbf{b} and \ref{fig:LowNA}\textbf{b}) the direct sum of all the low photon number images acquired with the 225 different SLM gratings which is similar to an incoherent image. Also, we show the low-resolution, coherent image obtained for a normal illumination in Fig. \ref{fig:Target2}\textbf{a} and \ref{fig:LowNA}\textbf{a}. It can be seen that the images reconstructed through FP have higher-resolution compared to the low-resolution images and they also show an improved resolution compared to the equivalent incoherent imaging scheme. The phase image of the wasp wing in Fig. \ref{fig:LowNA}\textbf{c} shows features that are not clear in the intensity image – the dark regions in the intensity image still contain phase information showing that the wing is not discontinuous.\\
Fig. \ref{fig:Target2}\textbf{c} is used to determine the resolution achieved. Using the slanted edge modulation transfer function method we find a resolution of 23 $\mu$m. The resolution achieved by the incoherent imaging is found to be of of 30 $\mu$m. In both cases the obtained field of view is of $\sim$2mm.\\
Finally, it is worth noting that that even-though the resolution and field of view of our optical system are here relatively limited, as observable on Fig. \ref{fig:Target2}, due to the technical choices we made here to implement the proposed technique, it should be noted that the technique imposes no inherent limit on the resolution of the system and that imaging/microscopy systems could be build that would harness our technique while achieving a much better resolution.

\section*{Discussion}

We have demonstrated the possibility of harnessing EPR-type quantum correlations to produce Fourier ptychographic (FP) images. We use FP combined with a heralded single-photon imaging scheme to allow the acquisition of images at extremely low light level and with a good signal to noise ratio. This is made possible by the implementation of a 'heralded imaging scheme' through the use of time-gated ICCD camera that temporally selects the heralded photons thereby removing parasitic contributions, e.g. dark noise, that would otherwise lead to a degradation of the image quality. By enabling the acquisition of low light level images with high signal to noise ratio, this promising technique allows the implementation of extremely low light level imaging in the context of microscopy. The proposed technique could therefore find applications in the context of delicate biological or chemical samples imaging. By enabling shot noise limited phase-amplitude imaging, our technique allows to image with a reduced exposure compared to technical noise limited classical technique. Moreover, further developments of our technique could exploit the sub-shot noise nature of the SPDC source to perform sub-shot noise imaging extracting both phase and amplitude about an object~\cite{brida2010experimental,moreau2017demonstrating}. It enables the extraction of both intensity and phase with only a limited number of detected photons with the help of Fourier ptychographic reconstruction. Finally, although the exact implementation reported in the present work relied on scanning the position of the heralding detector, it would be possible to implement a non scanning FP scheme based on a modification of the present scheme where the heralding photons would also be detected by a camera or a SPAD array placed in a Fourier plane of the object. The detection of correlations between different pixels of the two spatially resolved detectors would then enable a similar FP reconstruction as the one presented here to be performed but without the need for scanning and hence improving the overall efficiency.

\section*{Methods}
\subsection*{Experimental setup}
The principle of the image acquisition in this work is similar to the heralded imaging configuration reported in \cite{morris2015imaging}. The source is a 3 mm long $\beta$-Barium Borate (BBO) crystal, cut for type I phase matching, pumped by a quasi continuous laser at 355 nm. The parametric down-converted light is spectrally filtered to select photons at the degenerate frequency by an interference filter with a 10 nm bandwidth centred on 710 nm that is placed just after the BBO crystal (filter not shown on Fig. \ref{fig:setup}). The photon-pairs are then stochastically separated into a camera arm and a heralding arm by using a pellicle beam-splitter (BS). In the present configuration the object is placed in the far-field of the crystal output in the camera arm. It is illuminated by the signal (by convention) multi-mode SPDC beam with a full-width half-maximum of 156 \textmu{}m. The camera is positioned in an image plane of the object, with a magnification $M=\frac{4}{1.5}$ between the object plane and the camera image plane. In the heralding arm, a spatial light modulator (SLM) is placed in the far field of the crystal, i.e. in a plane equivalent to the object plane with identical magnification as the object plane. The idler photons reflected off the SLM are then collected into a single-mode fiber (SMF) using a fiber coupler set up such that retro-propagated light outputting the SMF input is collimated after the coupler. The output of the optical fiber is then coupled to the heralding SPAD, the output from which triggers the ICCD in order to acquire images heralded by the detection of an idler photon. Because the SMF fiber is collecting photons at a given position in a plane that corresponds to a Fourier plane of the object plane, the images acquired by the triggered ICCD correspond to images obtained by illuminating the object with the state described in equation (\ref{illumstate}). This is due the fact that when the camera is triggered conditionally on the detection of a photon by the SPAD, the state of the SPDC light is projected onto a post selected state for which the idler has a determined position. Note that the object is here positioned in the far field of the crystal and the SMF selects the transverse position of the idler photon $(x_f,y_f)$ that parametrise the angle of incidence of the illumination on the object $(\theta_x,\theta_y)$.\\
Changing the angular position of the fiber within the spatial extent of the idler can be implemented by changing the period of the blazed grating displayed on the SLM thereby changing the deflection angle of the idler beam. One can then perform a complete Fourier Ptychographic acquisition, by recording many such heralded images, each being acquired for a particular period and orientation of grating.
Each time the SPAD detects a photon, the camera intensifier is triggered with a gate width of a few nanoseconds. During this window, the camera single photons events hitting the intensifier are amplified and such events are then recorded by the CCD array. The intensifier can be triggered many times during the CCD chip exposure time such that the frames that are obtained are accumulations of all the single-photon events acquired during the exposure time.  The HI image is then formed from summing these individual frames, each one of which is photon-sparse.
There exists an electronic delay in the ICCD triggering that needs to be compensated to ensure that the photons detected in the images are from the same pair as the herald photons. This delay is achieved by introducing a 22 m image-preserving delay line in the camera arm as shown in Fig. \ref{fig:setup}.
The heralded images (as shown in Fig. 2a) correspond to 1000 frames each of 1 second of exposure, during which time the camera intensifier is triggered for every heralding detector pulse. The CCD sensor is air cooled to -30$^\circ$C. The images are thresholded to generate binary images that corresponds to detection of single photons. We can calculate the threshold over which a pixel is considered to correspond to a photo-detection and the noise probability per pixel by acquiring frames with the camera optical input blocked. The dark-count probability per pixel arising from the camera readout noise, is calculated to be around $5\cdot 10^{-5}$ per frame.
The images obtained through the DI method are acquired in a continuous trigger mode and we use the ICCD camera as a linear detector. Indeed, to retrieve a comparable number of photons acquired in comparison to the HI method, the total triggering time needs to be considerably longer for DI compared to HI as the photons are not heralded and are randomly detected by the camera. This results in most of the detected photons in DI being background noise events, resulting in a very poor signal to noise ratio compared to the HI (see\cite{morris2015imaging} for a comparison between HI and DI when both are used in single photon counting mode). For this reason, to obtain good quality DI image based on thresholded frames one would need to acquire a very high number of frames since each frame needs to be photon sparse and that most of the photo-detection are then noise. This would mean a very long acquisition to obtain the DI images. To avoid this extended measurement time we do not acquire the DI images in a single-photon counting mode and instead simply acquire the DI images by accumulating frames that are not photon sparse and use the ICCD as a linear detector by simply subtracting the strong background of the images evaluated with the camera optical input blocked.

To perform the FP reconstruction a set of $15\times 15$ images is acquired with an increment of $\Delta k_{\bot}\approx 4.2\ rad\cdot mm^{-1}$ in term of transverse wave vector for the plane wave illuminating the object, which corresponds to an increment $\Delta\theta\approx 5\cdot10^{-4} rad$ of the angle of incidence of the light. The magnification of the system from the SLM to the Camera is $M=\frac{4}{1.5}$, this means that for each increment of the angle $\Delta\theta$ corresponds to a tilt of $\Delta\theta'=\frac{\Delta\theta}{M}$, the mean direction of propagation of the post selected photons impinging on the camera. Such a tilt corresponds to a shift of 3 pixels in the Fourier domain obtained by proceeding to a Fast Fourier Transform on a 874 camera pixels region of interest.\\

\subsection*{Reconstruction procedure}
The reconstruction procedure starts by creating an initial estimate of the high-resolution image by interpolating the central low-resolution image (the high-resolution image contains more pixels due to the improved resolution). This interpolated image is Fourier transformed to obtain the high-resolution frequency spectrum of the object $O(\theta)$. Starting with this first evaluation, one chooses to work with one of the recorded illumination angles $\theta_i$, in our case this corresponds to a particular angular position of the single-mode fiber in the heralding beam. The complex spectrum $O(\theta)$ is then shifted according to the illumination angle $\theta_i$. In the subsequent step the complex spatial spectrum is filtered with the pupil function P of the imaging system. The resulting spectrum $O_{tmp}(\theta)=O(\theta-\theta_i)\times P(\theta)$ is propagated to the image plane through an inverse Fourier transform, where the image amplitude is then updated with the square root of the image $I_i$ recorded for the fiber position equivalent to the illumination angle $\theta_i$. Another Fourier transform is performed to propagate the resulting complex image back to the spatial frequency domain where the resulting new object spatial frequencies evaluation $O_{i}(\theta)$ is used to update the most current shifted reconstruction of the object $O(\theta-\theta_i)$. The result can then be shifted back by $\theta_i$ to obtained the new current evaluation of the object complex spectrum $O(\theta)$. The same steps are then performed again with the remainder of the images taken at different angles of illumination $I_i$, which completes one iteration of the reconstruction. The choice of the reconstruction method that is used determines the cost function used to update the object spatial frequencies using $O_{i}(\theta)$. As mentioned before, in order to minimise the impact of the strong shot-noise present in the images, we use a cost function corresponding to a sequential quasi Gauss-Newton optimisation method\cite{yeh2015experimental}. Similar cost function can also be used to update the pupil function to correct for any aberrations present in the system\cite{ou2014embedded}. The reconstruction process is iterated until the difference between the current reconstruction and the reconstruction from previous iteration falls below a predefined convergence criterion. The final high-resolution spatial frequency spectrum $O(\theta)$ is inverse Fourier transformed to obtain the high-resolution amplitude and phase reconstruction of the object.\\

\subsection*{Instrumentation}
To perform the experiment we used an Andor ICCD camera (Model iStar DH334T-18U-A3). The SPAD is an SPCM from Perkin-Elmer (SPCM-ARQ-13-FC). The SLM used to perform the scanning of the relative position of the beam and the fibre is an Hamamatsu LCOS-SLM (X10468). The Laser is a JDSU xCyte (Model CY-355-150).

\bibliography{sample}

\noindent 

\section*{Acknowledgements}
P-AM acknowledges the support from the European Union’s Horizon 2020 research and innovation programme under the Marie Sklodowska-Curie grant agreement No 706410, of the Leverhulme Trust through the Research Project Grant ECF-2018-634 and of the Lord Kelvin / Adam Smith Leadership Fellowship scheme. TA acknowledges support by the ‘EPSRC CDT in Intelligent sensing and Measurement'. PCK acknowledges support by the SUPA and the NC3Rs. This work was supported by the UK EPSRC (QuantIC EP/M01326X/1) and the ERC (TWISTS, Grant no. 192382).
The authors thanks Robert Boyd for helpful discussions concerning the realisation and Ermes Toninelli for helpful discussions concerning the experimental setup. 

\section*{Author contributions statement}

P-AM developed the original idea and conceived the experiment with contributions from MJP. P-AM conducted the experiment. TA and PCK developed the reconstruction algorithm under the supervision of ARH. TA and P-AM analysed the data. All authors contributed to the interpretation of the results and contributed to the manuscript.

\section*{Additional Information}
\subsection*{Competing interests :} The authors declare no competing interests.
\end{document}